\documentclass[usenatbib]{mn2e}

\usepackage{graphicx}
\graphicspath{{/Users/caz/Documents/PHD/idl/closepairs/ps/ps_to_eps/}} 
\usepackage{array}

\usepackage{amsmath,amssymb,mathrsfs}
\usepackage{fancybox,shadow}
\usepackage[titletoc]{appendix}
\usepackage[usenames]{color}
\usepackage{multirow}
\usepackage{dpfloat}
\usepackage{chngcntr}

\title[Star Formation and AGN Activity in Interacting Galaxies: A Near-UV Perspective]
  {Star Formation and AGN Activity in Interacting Galaxies: A Near-UV Perspective}
\author[C. Scott et al.]
  {Caroline Scott$^{1,2,3}$\thanks{email cscott@cfa.harvard.edu} and Sugata Kaviraj$^{4,5,1}$  \\
  $^1$Astrophysics, Imperial College London, Blackett Laboratory, London SW7 2AZ, UK\\
  $^2$Harvard-Smithsonian Center for Astrophysics, 60 Garden Street, Cambridge, MA 02138, USA\\
  $^3$Institute for Applied Computational Science, Harvard University, 29 Oxford Street, Cambridge, MA 02138, USA\\
  $^4$Centre for Astrophysics Research, University of Hertfordshire, College Lane, Hatfield, Herts, AL10 9AB, UK\\
  $^5$Department of Physics, University of Oxford, Keble Road, Oxford, OX1 3RH, UK\\}
\date{Accepted 18 October 2013}

\pagerange{\pageref{firstpage}--\pageref{lastpage}} \pubyear{2013}

\def\LaTeX{L\kern-.36em\raise.3ex\hbox{a}\kern-.15em
    T\kern-.1667em\lower.7ex\hbox{E}\kern-.125emX}

\newtheorem{theorem}{Theorem}[section]
{\begin{sfminipage}{0.9\textwidth}\begin{theorem}[#1]}
 {\end{theorem}\end{sfminipage}}

\makeatletter

\newcommand{\Rmnum}[1]{\expandafter\@slowromancap\romannumeral #1@}
\makeatother

\makeatletter
\@removefromreset{figure}{chapter}
\makeatother

\makeatletter
\renewcommand{\thefigure}{\@arabic\c@figure}
\makeatother

\begin{document}

\label{firstpage}

\maketitle

\begin{abstract}
We study nearby galaxies in close pairs to study the key factors affecting star formation and AGN activity triggered during galaxy interactions. Close pairs are selected from the Sloan Digital Sky Survey assuming a projected separation of $<$30kpc and recessional velocity difference $<$500km s$^{-1}$. Near-ultraviolet (NUV) fluxes from GALEX are used to estimate specific star formation rates (SSFRs). We find a factor of $\sim$5.3 increase in SSFR for low mass (10$^{8}$-10$^{11}$M$_{\odot}$) close pair galaxies and a factor of $\sim$2.1 increase in SSFR for high mass mass (10$^{11}$-10$^{13}$M$_{\odot}$) close pairs compared to the general galaxy population. Considering galaxies of all masses, we find a factor of $\sim$1.8 enhancement in SSFR for close pairs in field environments compared to non-pairs, with no significant increase for pairs in group and cluster environments. A modest decrease of a factor of $\sim$1.4 is found in the Seyfert fraction in close pair galaxies when compared to isolated galaxies, which suggests that mergers may not trigger AGN activity at the close-pair stage or may trigger a different class of AGN. This becomes a factor of $\sim$4.2 decrease when we restrict our analysis to high mass close pairs in group or cluster environments. 
\end{abstract}


\section{Introduction}

Mergers are fundamental to the standard hierarchical paradigm of galaxy formation and evolution. Models predict that they produce intense star formation episodes \citep{Mihos1996,Cox2008,Bournaud2011}, contributing to the build-up of stellar mass and black holes \citep{Sanders1996,Hopkins2005,DeBuhr2011}, and alter the morphological mix of the Universe \citep{Toomre1972,Kauffmann1993,Mihos1995,Kauffmann2000}. Past studies from the previous generation of astrophysical instruments found that interactions could enhance star formation in close pair galaxies \citep{Larson1978,Joseph1984,Lonsdale1984} and potentially trigger AGN activity \citep{Smith1986,Hutchings1987,Sanders1988}. Since the emergence of large-scale galaxy surveys, such as the Sloan Digital Sky Survey (SDSS) and the Galaxy Evolution Explorer (GALEX), robust statistical analyses can now be conducted on galaxy-galaxy interactions, resulting in extensive studies of mergers over recent years.

Substantial observational data now serves to complement results from simulations, which claim that close pair interactions and mergers cause the instability needed for gas clouds to collapse and produce starbursts \citep[e.g.][]{Kauffmann2000}. Spectroscopic and photometric evidence from large scale surveys such as IRAS \citep{Kennicutt1987} and the SDSS \citep{Ellison2008}, and high resolution instruments like the HST \citep{Patton2005} have bolstered such claims. Morphological asymmetry effects are also a trade mark signature of close pair systems; with most of the galaxies in the Arp catalogue \citep[Atlas of Peculiar Galaxies,][]{Arp1966} showing signs of recent tidal interactions \citep{Larson1978} and $\sim$40\% of close pair systems expected to show asymmetry effects \citep{Patton2005}. Interaction-induced effects are expected to depend on properties of the progenitor galaxies and external pair-properties of close pair systems. Such properties may include the environment in which the merger is taking place, pair separation, the type of galaxies merging (e.g.\ morphology, mass), and central galactic processes within the progenitors; such as feedback from AGN activity.

Close pair systems which are in the early stages of merging generally show less resulting star formation than further advanced mergers \citep[e.g.][]{Larson1978,Barton2000}. Using H$\alpha$ as a diagnostic for star formation for a large sample of spectroscopically classified SDSS DR5 close pairs, \cite{Woods2007} found evidence that the lower mass progenitor in a minor merger will experience the most star formation. \cite{Ellison2008} found an enhancement in star formation in close pairs of up to $70\%$ compared with a control sample of 40,095 SDSS galaxies which had the same mass distribution. This enhancement in star formation was greatest for major mergers; where the progenitors have mass ratio M$_{1}$/M$_{2} > 0.5$. \cite{Wong2011} used UV photometry from GALEX to look at NUV-r and FUV-r colours for intermediate redshift close pairs ($0.25\leq z \leq0.75$) drawn from the Prism Multi-Object Survey (PRIMUS). They find an $\sim$$15-20\%$ increase in SSFR for close pairs with projected separation $\leq$50$h^{-1}$ kpc, and an $\sim$$25-30\%$ increase in SSFR for close pairs with projected separation $\leq$30$h^{-1}$ kpc.

Dense environments, such as galaxy clusters, often have lower gas fractions as a result of tidal fields and ram pressure stripping \citep{Byrd1990}; as a consequence, less gas is thought to be available to fuel star formation during a merger. Accordingly, close pairs in galaxy clusters and groups have been found to show comparatively less star formation than close pairs in the field \citep{Kauffmann2004,Alonso2005,Ellison2010}. Using SDSS DR4 data, \cite{Ellison2010} found a decrease in star formation in local dense environments when using asymmetry and optical colours to indicate interaction-induced star formation. They concluded that the higher levels of star formation detected in close pairs in low density environments is a result of the higher gas fraction available to fuel star formation. 

Studies of radio and QSO galaxies found that a significant number of radio galaxies are currently interacting or show recent signs of merger actvity \citep[e.g.][]{Smith1986,Hutchings1987}; suggesting a link between close pair galaxies and AGN activity. \cite{Hutchings1987} noted that AGN activity is generally observed in the larger of two interacting galaxies and suggested that in minor mergers the smaller galaxy can serve to fuel nuclear activity in the massive (often elliptical) galaxy.

Since recently formed stars  ($\lesssim$1Gyr) are responsible for most of a galaxy's UV luminosity, recent star formation rates strongly correlate with UV luminosity \citep{Iglesias-Paramo2006}. \cite{Salim2005} show that NUV-$r$ colour alone is sufficient to estimate the star formation history of galaxies and that GALEX is sensitive to star formation levels as low as $\sim$10$^{-3}$~M$_{\odot}$ yr$^{-1}$. GALEX's NUV band (effective wavelength: 2271\AA) allows us to study star formation with much more sensitivity to recent star formation than optical filters \citep{Yi2005,Kaviraj2007,Donas2007,Bianchi2011}. 

Emission line tracers of star formation (e.g. H$\alpha$) often limit measurements to the central galactic regions due to finite fibre size. SDSS fibres have a size of 3'' and therefore they measure a different fraction of light in each galaxy. Even though corrections can be made based on the ratio of the Petrosian-to-fibre flux, this correction is somewhat uncertain \citep[e.g.][]{Brinchmann2004}. However, UV photometry provides a measure of emission from young stars for the full extent of a galaxy. One caveat is the NUV band is extremely sensitive to interstellar reddening, more-so than optical and IR photometry \citep{Meurer1999,Pannella2009}. We correct our GALEX flux measurements for interstellar extinction using the Balmer decrement from the SDSS spectrum.

In this work recent star formation is examined in close pair galaxies at various stages of the merger process. Properties such as mass, close pair (projected) separation, environment and AGN activity are investigated to see how they affect star formation in pairs; where possible we try to break degeneracies between these properties. This work offers a similar investigation to some previous research, but from an optical-UV perspective and with a broad range of derived properties to investigate close pairs with various characteristics in detail. 

In Section 2 we describe how the close pairs and wide pairs samples were extracted from the SDSS database, the process by which these samples were cross-matched with GALEX data and how the properties for our study (such as mass, environment, BPT classification etc.) were derived. In Section 3 we present the results of our analysis. In Section 4 we look at the fraction of Transition, LINER and Seyfert galaxies in projected separation bins from 0-150kpc to investigate potential changes to these fractions as merging galaxies draw closer together. The evolution of these fractions with decreasing projected separation is studied as a function of mass and environment. In Section 5 the results are summarised and discussed.

\section{Sample Description}
\subsection{SDSS and GALEX}\label{sdsssection}

Our close pairs catalogue is extracted from the SDSS Data Release 7 (DR7) database \citep{Fukugita1996,Gunn1998,York2000}. The SDSS uses a multi-object fibre spectrograph to observe spectra over an area of $\sim$10,000 deg$^{2}$. Photometric data is imaged using five optical filters $u,g,r,i,z$ over a 3,000-11,000\AA\ range (where 3,000\AA\ is the atmospheric UV cut-off wavelength and 11,000\AA\ is the silicon sensitivity limit of the CCDs) using a large format mosaic CCD camera. Spectra for objects within $55''$ can only be obtained if they are observed in overlapping tiles.  These so-called \emph{fibre collisions} present an incompleteness issue when studying galaxy close pairs as only $\sim$30\% of the $10,000$ deg$^{2}$ area covered by the SDSS is observed by overlapping tiles \citep{Darg2010}. However, since the pairs that are detected are drawn randomly from a homogenous sample, our catalogue constitutes a representative (yet incomplete) sample of low redshift close pairs.

GALEX (Galaxy Evolution Explorer) is a NASA space based all-sky survey observing at ultraviolet wavelengths with 4-6$''$ resolution and $\sim$50cm$^{2}$ effective area \citep{Morrissey2007}. GALEX images in $1_{.}^{\circ}2$ diameter circular fields at 1770-2730\AA\ (NUV) and 1350-1780\AA\ (FUV) simultaneously, using a modified Ritchey Chr\'{e}tien telescope. The FUV limit is 5000 cts/s, m$_{AB}=9.5$, $F_{\lambda} = 7\times10^{-12}$ erg cm$^{-2}$ s$^{-1}$\AA$^{-1}$ and the NUV limit is 30,000 cts/s, m$_{AB}=8.9$, $F_{\lambda} = 6\times10^{-12}$ erg cm$^{-2}$ s$^{-1} $\AA$^{-1}$. 

We use the GR4/GR5 database which combines data from the following imaging surveys: All-sky Imaging Survey (AIS), Deep Imaging Survey (DIS), Medium Imaging Survey (MIS), Nearby Galaxy Survey (NGS) \citep{Martin2005,Morrissey2005}. Our sample is mainly from the AIS and MIS surveys. The AIS has 100s exposure time, 26,000deg$^{2}$ sky coverage, m$_{AB}=20.5$ depth and 28,000 tiles. The MIS has 1,500s exposure time, 1,000deg$^{2}$ sky coverage, m$_{AB}=23.5$ depth and 1615 tiles. AIS covers around 3/4 of the sky and aims to provide an all-sky survey with similar depth to the Palomar Observatory Sky Survey \Rmnum{2} and the SDSS. Regions in the vicinity of the Galactic plane and the Magellanic clouds were avoided due to the sensitivity of detectors in order to safeguard the detectors from potentially over-saturating UV fluxes. As a result, imaging in the surrounding areas can be patchy, but otherwise fields are generally adjacent. MIS is also positioned to have a significant overlap with the SDSS; there is a 7325 deg$^{2}$ overlap between SDSS DR7 and AIS (GR5) and 1103 deg$^{2}$ with the MIS (GR5) \citep{Bianchi2011}.

\subsection{Pair Extraction and Cross-Matching}

To extract close pairs from the SDSS DR7 database we use a procedure to seek galaxies with small angular separation and a small recessional velocity difference (i.e. low separation in the line-of-sight direction). We follow \cite{Patton2002} who suggest a projected separation of 20$h^{-1}$kpc and a difference in recessional velocities of 500km s$^{-1}$. \cite{Soares2007} compares projected separation with spatial separation for a Monte Carlo simulated sample of gravitationally-bound pairs and finds that over 50\% with projected separation $\leq50$kpc actually have spatial separation $>50$kpc. Thus, we accept that projected separation is merely an approximation to spatial separation, and that the physical separation is likely to be higher. Incompleteness due to fibre collisions, large peculiar velocities, minor mergers (low mass galaxies are often below the SDSS spectroscopic limit of $r<17.77$) etc.\ result in only $\sim$30\% of close pairs being detected. Since the sample used is from a low redshift survey and is magnitude limited, we do not impose a maximum redshift constraint. We remove all systems where more than one SDSS object (including the primary object) is within 5'' of a GALEX object (5'' being the GALEX resolution). The median redshift is z$\sim$0.07. The close pairs sample consists of 6668 galaxies with SDSS data and 2902 galaxies with both SDSS and NUV photometry. A \emph{wide pairs} sample with projected separation 30-150kpc and $\Delta z \sim 0.0017$ was also extracted to be used as a control sample with which to compare the close pairs sample. The wide pairs sample consists of 34,294 galaxies with SDSS data and 19,202 galaxies with both SDSS and NUV data.

\begin{figure*}
\begin{center}
\begin{tabular}{cc}
\includegraphics[width=0.48\textwidth]{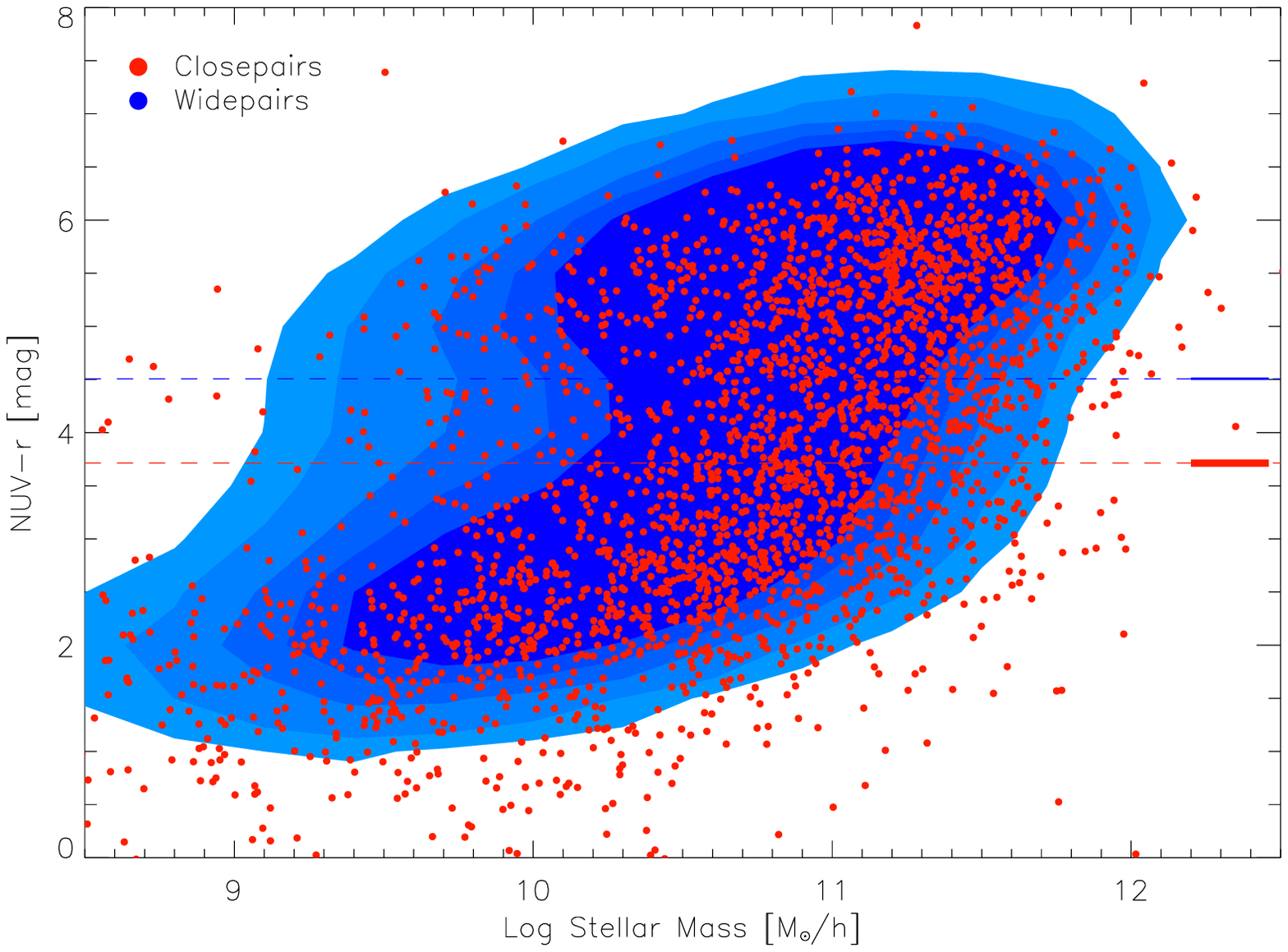} &
\includegraphics[width=0.48\textwidth]{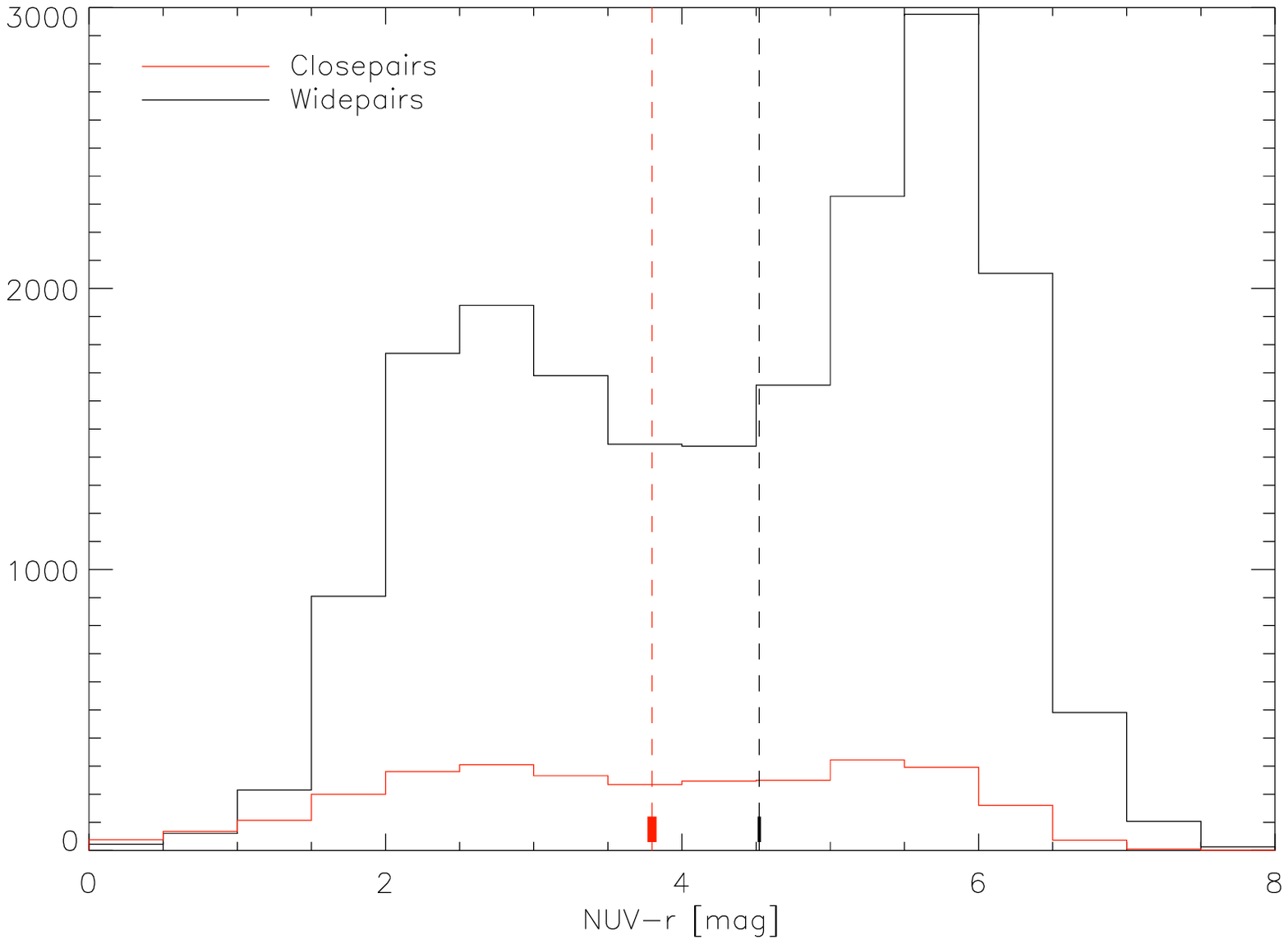} \\
\includegraphics[width=0.48\textwidth]{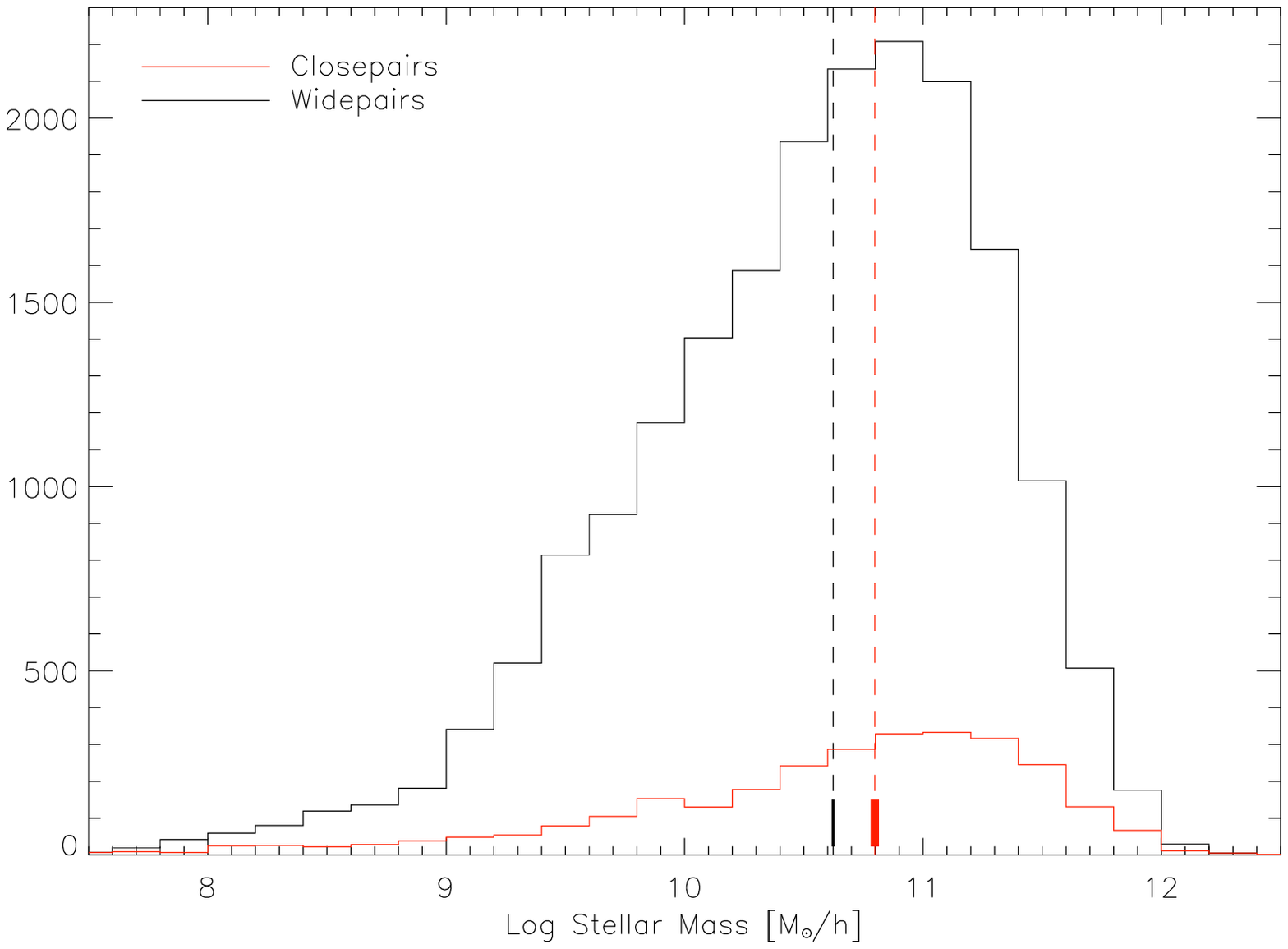} &
\includegraphics[width=0.48\textwidth]{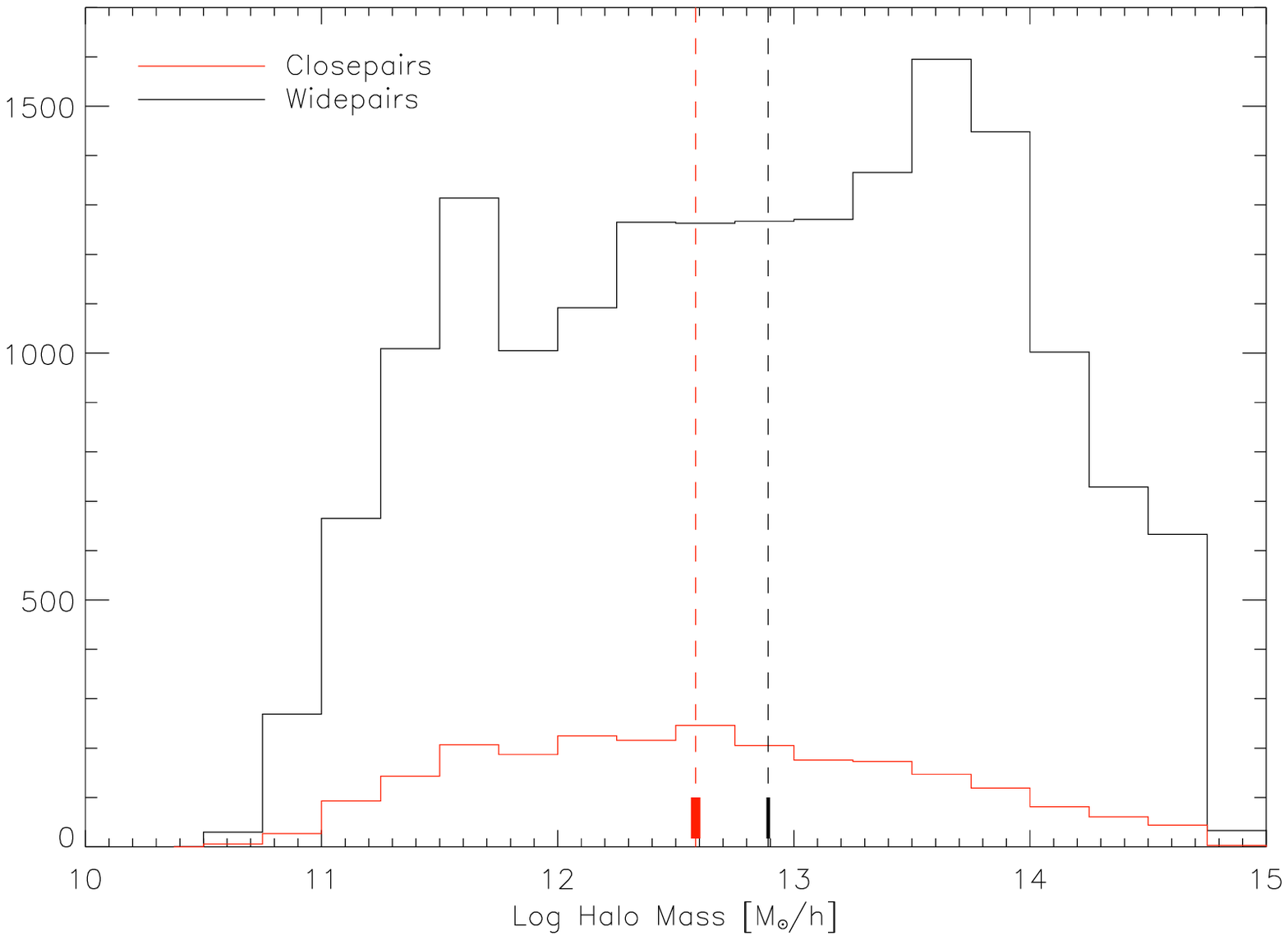}
\end{tabular}
\end{center}
\caption{\small{Top: NUV-$r$ against stellar mass plot (left) and histogram (right) for the wide pairs (black) and close pairs (red) samples. Bottom: stellar mass distribution (left) and halo mass distribution (right) for the wide and close pairs samples. Dotted lines represent median values, and coloured blocks on median lines show bootstrapping results.}}\label{cm_mass_plots}
\end{figure*}

Kcorrect V4\_2 \citep{Blanton2007} was used to calculate K-corrections for the (galactic extinction-corrected) SDSS/GALEX apparent magnitudes, then absolute magnitudes were derived. Stellar masses were approximated using the following formula from \cite{Wang2006} (based on \citealt{Bell2003});
\begin{align}
\textrm{log (M}_{\ast}/\textrm{M}_{\odot}) ={}& -0.4[\textrm{Mag($r$)}-4.67]-0.306 \notag \\
{}&+1.097[\textrm{Mag($g$)}- \textrm{Mag($r$)}]+0.15, \label{mass_eqn}
\end{align}where 4.67 is the $r$-band solar absolute magnitude from the SDSS. This assumes a \cite{Salpeter1955} stellar IMF with dN/dM $\propto$ M$^{-2:35}$ and 0.1M$_{\odot} < $ M $ <100 $M$_{\odot}$. The median uncertainty in stellar mass values is $\sim$0.1 dex, with maximum uncertainty expected to be $\sim$0.2 dex.

Figure \ref{cm_mass_plots} (top) shows the NUV-$r$ distribution for the close and wide pairs samples. The bottom left histogram shows the stellar mass distribution; the close pairs sample has a median stellar mass of 10$^{10.8}$M$_{\odot}$ and the wide pairs sample has a slightly lower median stellar mass of 10$^{10.6}$M$_{\odot}$. \cite{Bundy2009} and \cite{Darg2010} find a similar bias towards higher stellar masses (by $\sim$0.2 dex) for close pairs. This was suggested to be a result of the increased merger fraction in higher density environments, where massive red spheroidal galaxies are more common. However, \cite{Simard2011} show that photometry for close pairs from the standard SDSS pipeline is sometimes poor for pairs with projected separation $\lesssim$20kpc and \cite{Patton2011} suggest that this has led some authors to incorrectly perceive an extremely red population in close pairs samples. This reddening effect would explain our bias towards higher stellar masses for the close pairs, however, since the difference lies within the stellar mass error we justify their use.

\subsection{Environment and Emission-Line Analysis}

The close and wide pairs samples were crossmatched with an environment catalogue derived from a halo-based group finder \citep{Yang2005,Yang2007}. This method primarily uses the friends-of-friends algorithm to find groups \citep{Huchra1982}, then approximates the group centre using its brightest member. An initial mass is calculated using the mass-to-light ratio, then the total luminosity of potential groups is estimated using the luminosity function from \cite{Norberg2002}. All galaxies are then assigned a probability of belonging to each group's dark matter halo. The algorithm iteratively assigns each galaxy to its most probable group (merging smaller groups that can be identified as one), updates the assigned group centre, and re-calculates the total luminosity. The dark matter halo mass is then calculated from this finalised characteristic luminosity, and is used to approximate environment density. Figure \ref{cm_mass_plots} (bottom right) shows the distribution of the halo mass parameter for both samples. We assume that a halo mass value from 10$^{10}$ to 10$^{13}$M$_{\odot}$ describes a local field environment, 10$^{13}$ to 10$^{14}$M$_{\odot}$ a group environment, and 10$^{14}$ to 10$^{15}$M$_{\odot}$ a cluster environment \citep{Kaviraj2009}.

\begin{figure*}
\begin{flushleft}
\begin{tabular}{cc}
\includegraphics[width=0.48\textwidth]{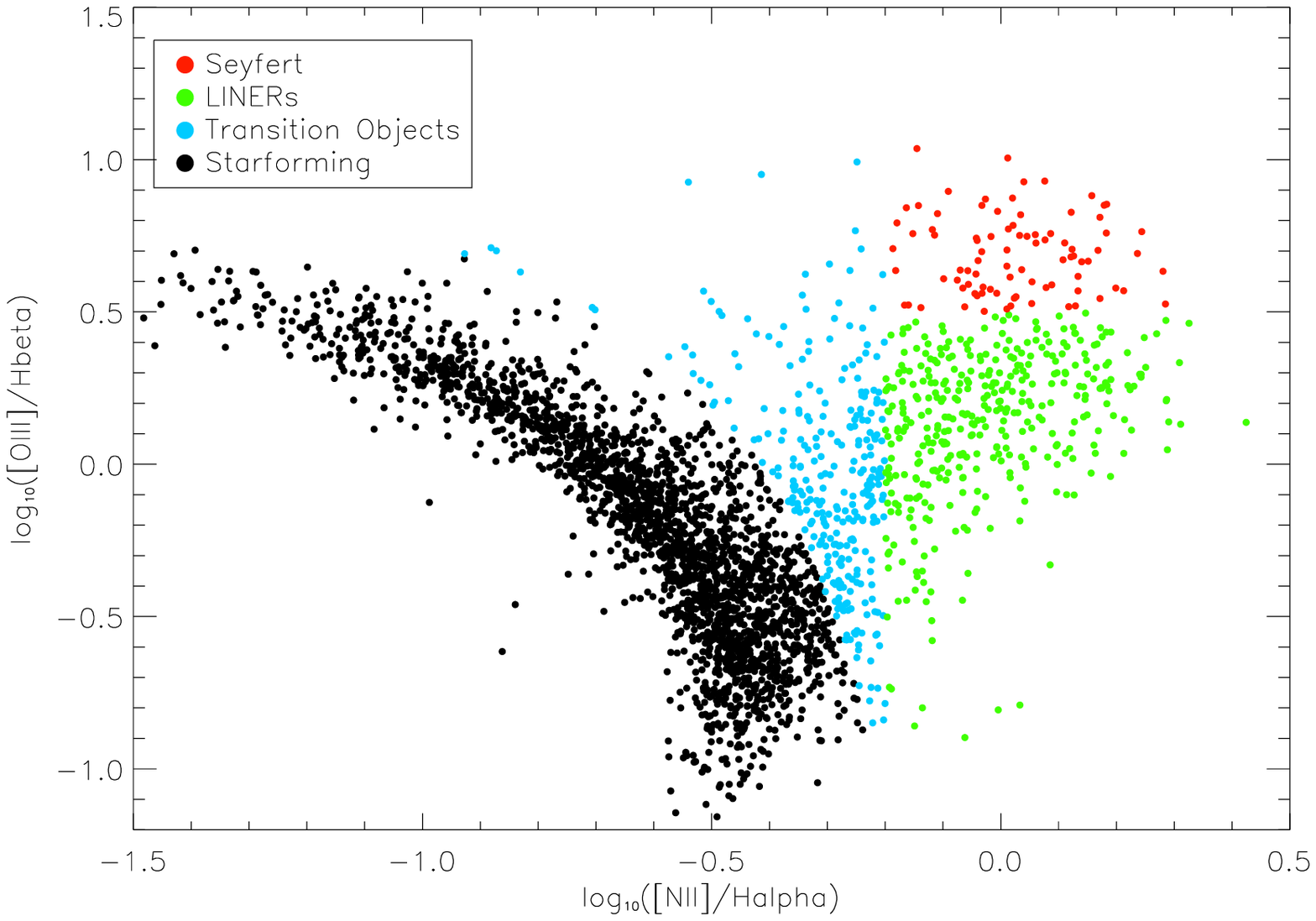} &
\includegraphics[width=0.48\textwidth]{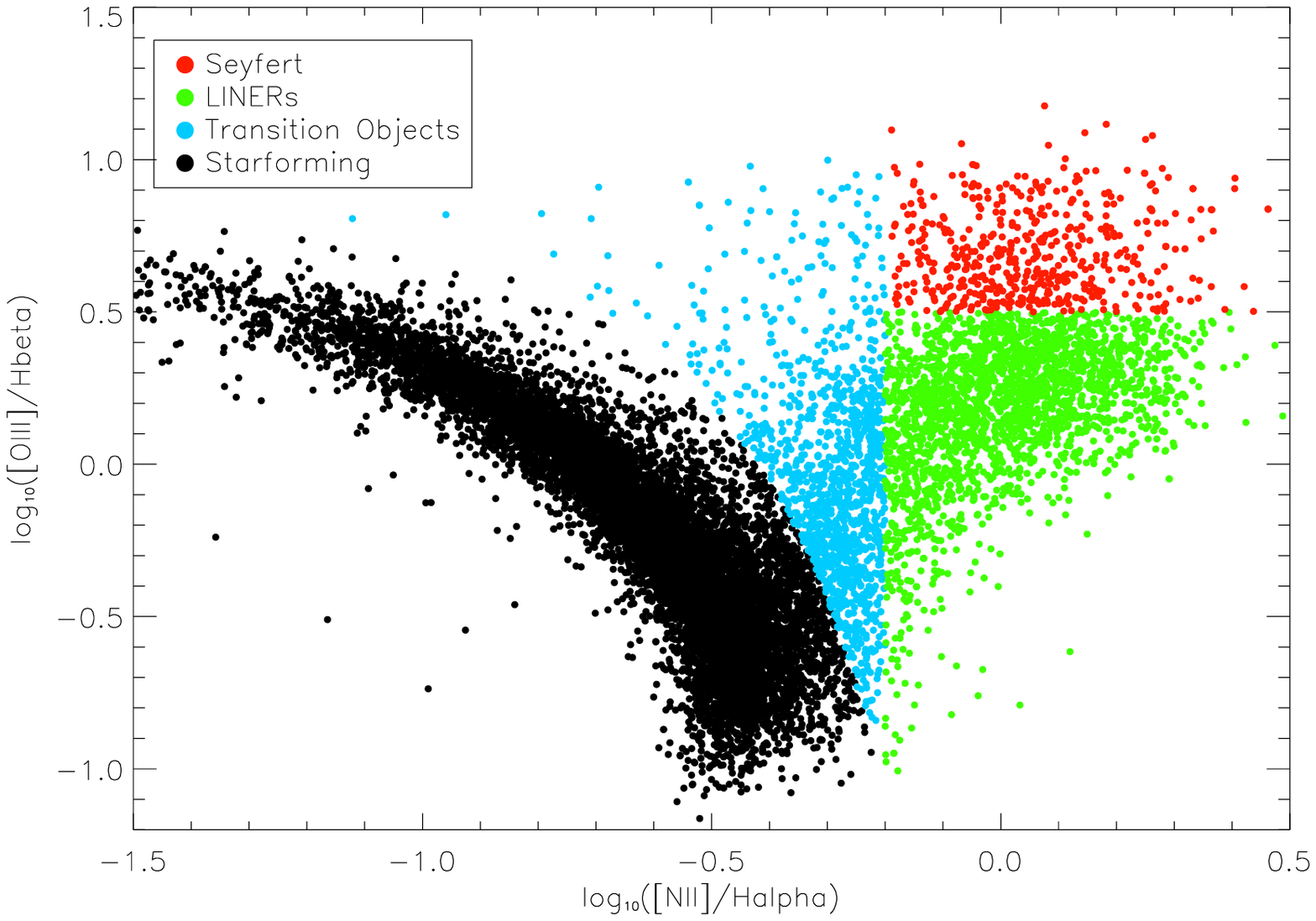}
 \end{tabular}
\caption{\small{BPT: log$_{10}$([N\Rmnum{2}]/H$\alpha$) is plotted against log$_{10}$([O\Rmnum{3}]/H$\beta$) for the close pairs (left) and wide pairs (right). Objects are classified as Starburst, Transition, LINER or Seyfert depending on the region they inhabit on the BPT diagram.}}\label{bptplot}
\end{flushleft}
\end{figure*}

A BPT analysis \citep{Baldwin1981,Kewley2001,Kauffmann2003,Kewley2006,Trichas2010,Kalfountzou2011} allows us to determine the predominant mechanism of excitation in our sample galaxies. Intensity ratios of pairs of strong emission lines are used to separate Starburst galaxies from AGN. In Figure \ref{bptplot}, we plot log([N\Rmnum{2}]/H$\alpha$) against log([O\Rmnum{3}]/H$\beta$) and classify each galaxy as Starburst, Transition, LINER or Seyfert depending on where they lie on the BPT diagram; as defined by \cite{Kauffmann2003}. A Quiescent class is defined, though very few of our galaxies lie in the Quiescent region. Type \Rmnum{1} (i.e. unobscured) AGN are flagged and removed before the anaylsis. We consider LINER and Seyfert galaxies as Type \Rmnum{2} AGN. Only galaxies with [N\Rmnum{2}] and H$\alpha$ emission line signal-to-noise ratio $>$3 are used.

\subsection{Deriving Star Formation Rates}

We compute star formation rates (SFRs) for galaxies in our sample from their NUV absolute magnitudes using the following expression from \cite{Iglesias-Paramo2006}:
\begin{equation}
\textrm{log SFR}_{NUV}(\textrm{M}_{\odot}\textrm{yr}^{-1}) = \textrm{log }\textrm{L}_{NUV}(\textrm{L}_{\odot})-9.33\label{sfr}
\end{equation}NUV fluxes were first corrected for internal reddening in the galaxy via the Balmer decrement, measured using the Gas and Absorption Line Fitting code (GANDALF; \cite{Sarzi2006}). GANDALF fits stellar population and emission line templates to the galaxy spectrum simultaneously, to separate the stellar continuum and absorption lines from the ionized gas emission. It calculates an internal E(B-V) from the emission lines, in the standard way, via the Balmer decrement assuming Case B recombination. This internal E(B-V) likely traces the E(B-V) in the star-forming regions and is used to derive intrinsic NUV fluxes and star formation rates. To correct for internal and galactic extinction, we multiply the E(B-V) component by 2.751 (according to \citealt{Schlegel1998}) to correct our SDSS r-band magnitudes and by 8.2 (according to \citealt{Calzetti2000,Kaviraj2007}) to correct our GALEX NUV magnitudes.

\begin{figure}
\begin{flushleft}
\includegraphics[width=0.48\textwidth]{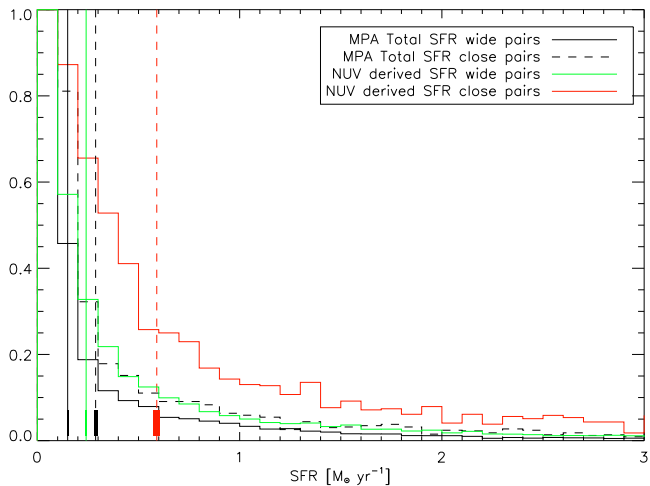} 
\end{flushleft}
\caption{\small{Emission line luminosity-derived SFRs from the MPA SDSS DR7 catalogue (black) are compared with NUV photometry-derived SFRs for our wide pairs sample (green) and close pairs sample (red). Vertical lines show sample median values.}}\label{mpa_sfr}
\end{figure}

We compare our NUV SFRs with those derived via H$\alpha$ fluxes in the MPA-JHU SDSS DR7 catalogues for the same galaxies\footnote{www.mpa-garching.mpg.de/SDSS/DR7/}. These were derived from emission line luminosities (based on \citealt{Brinchmann2004}). Our wide pairs sample provides a suitable comparison between SFR measurements for the general galaxy population (see Figure \ref{mpa_sfr}). For the wide pairs, our NUV-derived SFRs have a slightly higher SFR distribution than the MPA emission line derived SFRs but overall show good agreement; whereas for close pairs the SFR is generally higher when measured using NUV luminosities. Note that the total emission line SFRs are calculated by scaling the H$\alpha$ flux by the ratio of the fibre and Petrosian fluxes. The reason for the discrepancy could lie in this scaling, which may not fully capture the age gradient in the galaxy. We divide the NUV photometry-derived SFRs by the stellar mass attained from \eqref{mass_eqn} for each galaxy to get specific star formation rates (SSFRs). 

\section{Results}

\subsection{Enhancement of Star Formation as a Function of Separation and Mass}\label{sep_and_mass}

\begin{figure}
\centering
\includegraphics[width=0.48\textwidth]{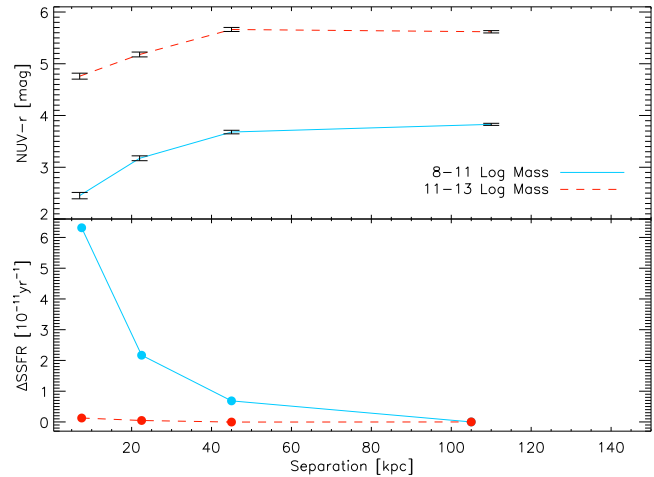}
\caption{\small{Top: Median NUV-$r$ colours for close and wide pairs binned by separation (0-15kpc, 15-30kpc, 30-60kpc and 90-130kpc) for both low stellar mass galaxies (10$^{8}$-10$^{11}$M$_{\odot}$, shown in blue) and high stellar mass galaxies (10$^{11}$-10$^{13}$M$_{\odot}$, shown in red). Bottom: SSFR difference between the separation bin in question and the widest separation bin, for both low and high mass galaxies. The fractional $\Delta$SSFR error is $\sim$10$\%$ for the low mass sample and $\sim$30$\%$ for the high mass sample.}}\label{median_plot_mass}
\end{figure}

\begin{table*}
\centering
\begin{tabular}{lllll}
 \hline
\textbf{Mass Range (Log$_{10}$M$_{\odot}$)} &   &   \multicolumn{2}{c}{\textbf{Projected Separation (kpc)}} & 	 \\
	&   0-15   & 15-30 & 30-60  & 90-130 \vspace{2 mm}
\\
\hline
\hline 
8-11       & 8.24$\times10^{-11}$     ~~~~~~    & 3.77$\times10^{-11}$   ~~~~~~       &  2.63$\times10^{-11}$  ~~~~~~~ & 1.94$\times10^{-11}$  ~~~~~~ \\
\hline
11-13     & 3.22$\times10^{-12}$ &  1.98$\times10^{-12}$  & 1.22$\times10^{-12}$ 	& 1.31$\times10^{-12}$  \\
\hline
\end{tabular}
\caption{\small{Median SSFR (yr$^{-1}$) derived from NUV luminosity for each stellar mass and separation bin.}}\label{mass_ssfr_table}
\end{table*}

In Figure \ref{median_plot_mass} we split the close pairs and wide pairs samples into low stellar mass (10$^{8}$-10$^{11}$M$_{\odot}$, shown in blue) and high stellar mass (10$^{11}$-10$^{13}$M$_{\odot}$, shown in red) galaxies. To investigate the star formation enhancement in advancing mergers, both low mass and high mass galaxies are binned further by projected separation and we analyse median NUV-$r$ trends in these bins as separation decreases. For both stellar mass bins, the median NUV-$r$ colours become bluer as projected separation decreases, indicating that recent star formation is enhanced as pairs draw closer together. This decrease occurs at approximately the same rate for the lower and the higher stellar mass bins. 

In Table \ref{mass_ssfr_table} we show the median SSFR (yr$^{-1}$) for the galaxies in each mass/separation bin. We also plot the difference in SSFR between the widest separation bin and the separation bin in question (this quantity is by definition zero for the widest separation bin). We find a difference of 6.3$\times10^{-11}$yr$^{-1}$ in SSFR from the widest (90-130kpc) to the smallest separation bin (0-15kpc) for low stellar mass galaxies, and a difference of 1.3$\times10^{-12}$yr$^{-1}$ for high stellar mass galaxies. Treating the widest separation bin as a control sample, representative of the general population of non-close pair galaxies, this indicates a factor of 5.3 increase in SSFR for low stellar mass close pair galaxies and a factor of 2.1 increase in SSFR for high stellar mass close pairs compared to the general galaxy population.

\subsection{Enhancement of Star Formation as a Function of Separation and Environment}\label{sep_and_env}

\begin{figure}
\begin{center}
\includegraphics[width=0.5\textwidth]{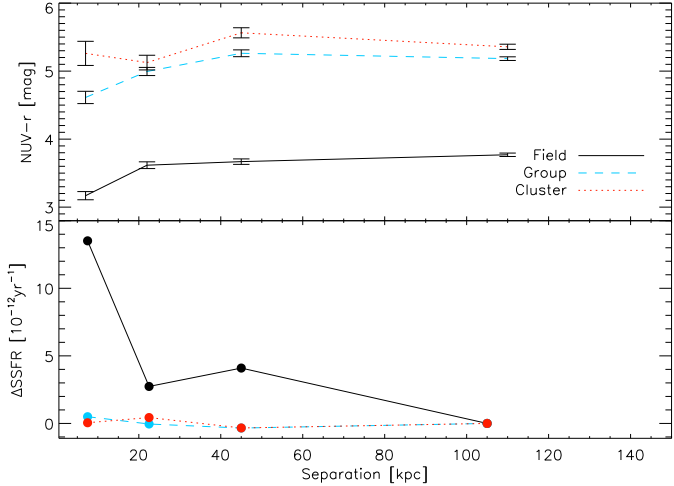}
\caption{\small{Top: Median NUV-$r$ colours for close and wide pairs binned by separation (0-15kpc, 15-30kpc, 30-60kpc and 90-130kpc) for pairs in field (black), group (blue) and cluster (red) environments. Bottom: SSFR difference between the separation bin in question and the widest separation bin, for each environment. The fractional $\Delta$SSFR error is $\sim$20$\%$.}}\label{median_plot_env} 
\end{center}
\end{figure}

\begin{figure*}
\begin{flushleft}
\begin{tabular}{cc}
\includegraphics[width=0.48\textwidth]{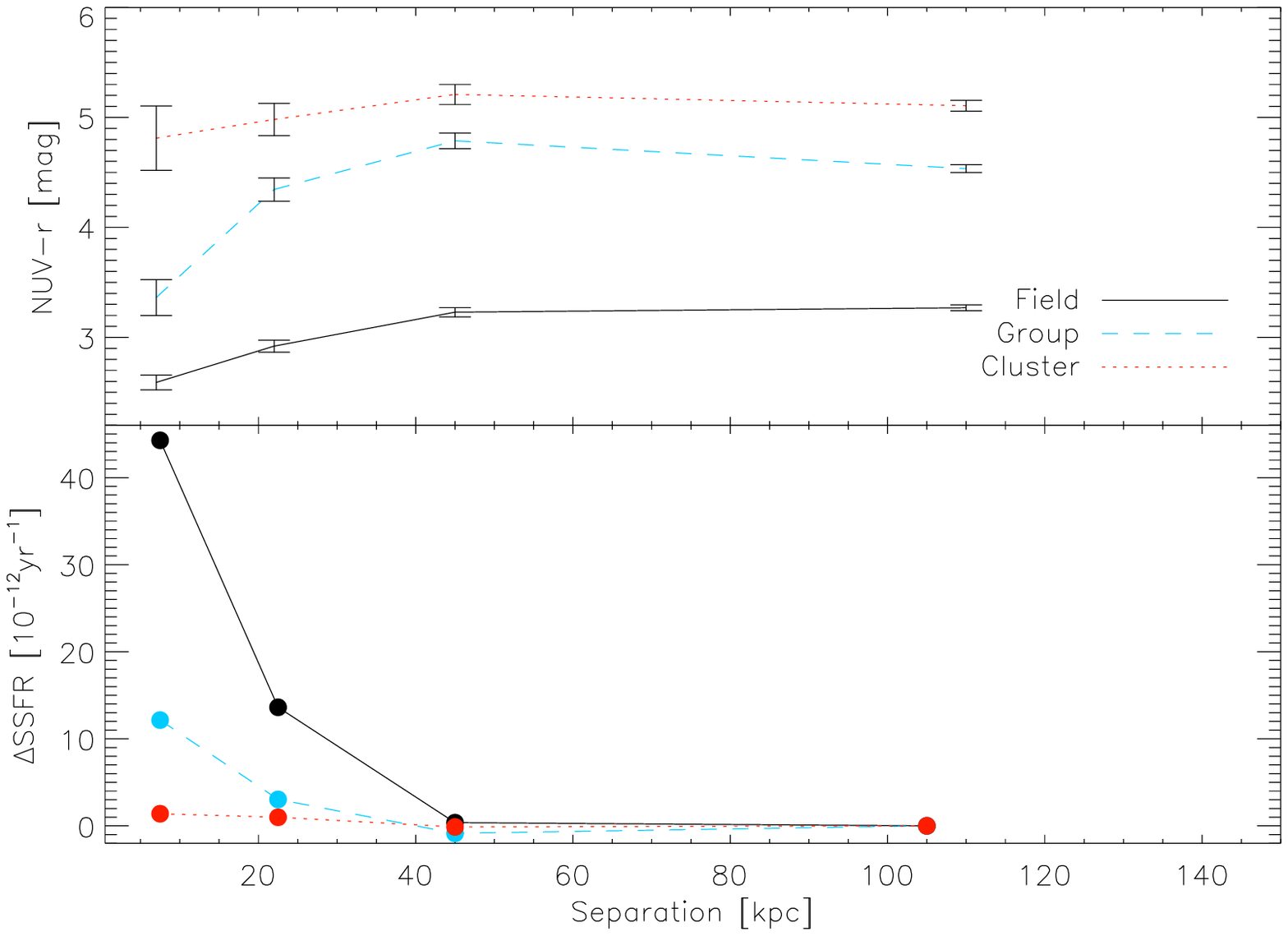} &
\includegraphics[width=0.48\textwidth]{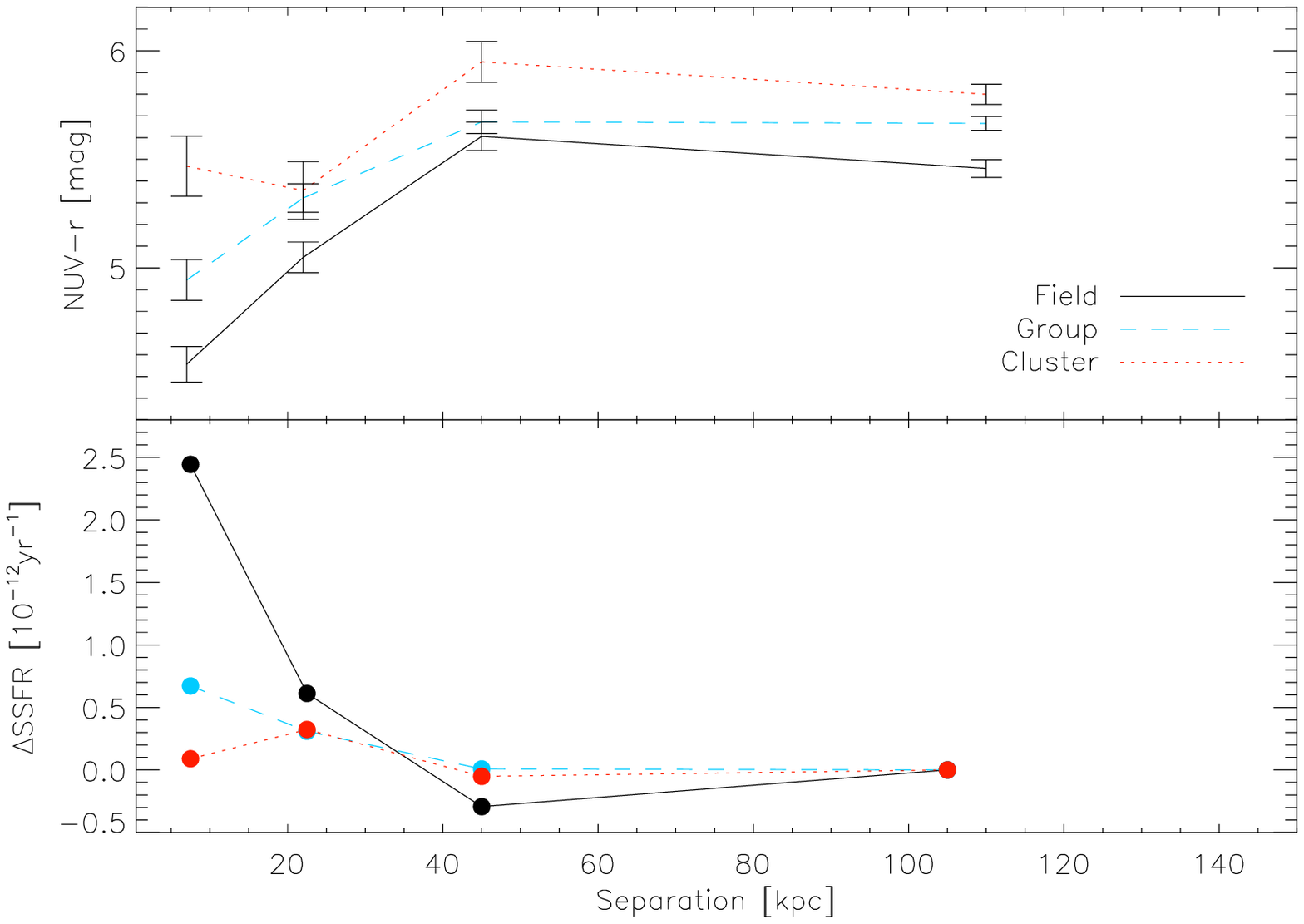}
 \end{tabular}
\caption{\small{Median NUV-$r$ values for each environment/separation bin are plotted for low stellar mass galaxies (10$^{8}$-10$^{11}$M$_{\odot}$ -left) and high stellar mass galaxies (10$^{11}$-10$^{13}$M$_{\odot}$ -right).}}\label{median_plots_env_low_high_mass}
\end{flushleft}
\end{figure*}

We now split the sample by environment and separation instead of stellar mass and separation. The three environment bins are field, group and cluster. We see in Figure \ref{median_plot_env} that the NUV-$r$ colour distribution shifts to bluer colours from the high density cluster environment to the low density field environment. As well as for general population galaxies in lower density environments, studies show strong evidence that particularly enhanced star formation takes place in close pairs in lower density environments \citep{Kauffmann2004,Alonso2005,Ellison2010}. We now quantify this from an NUV perspective.

We find a difference of 1.4$\times10^{-11}$yr$^{-1}$ (i.e. a factor of 1.8 increase) in SSFR from the widest to the smallest separation bin for pairs in field environments, and no significant increase for pairs in group and cluster environments. From our NUV perspective, this bolsters claims that decreasing local environment density leads to an increase in interaction-induced star formation for close pair galaxies which is greater than the enhancement expected for general population galaxies in low density environments.

Since stellar mass and environment are correlated, such that increasingly massive galaxies tend to be found in higher density environments, we attempt to break the degeneracy by splitting our sample by stellar mass and environment. Figure \ref{median_plots_env_low_high_mass} shows the environment/separation analysis, but now restricted to low mass galaxies (10$^{8}$-10$^{11}$M$_{\odot}$ -left) and higher mass galaxies (10$^{11}$-10$^{13}$M$_{\odot}$ -right). In the low mass analysis, NUV-$r$ colours in field and group environments become noticeably bluer with decreasing projected separation; with no clear trend for cluster environments. We find a rise in SSFR of 4.4$\times10^{-11}$yr$^{-1}$ (a factor of 2.4 increase) for field pairs and a rise of 1.2$\times10^{-11}$yr$^{-1}$ (a factor of 3.4 increase) for group pairs in the low mass sample. For high mass pairs the NUV-$r$ range is much narrower and lies in the redder colour region, implying that less recent star formation is being triggered. We see a rise of 3.9$\times10^{-12}$yr$^{-1}$ (a factor of 2.5 increase) in SSFR for high mass pairs in field environments.

\section{AGN activity in Close Pairs}

We wish to study how AGN activity evolves as a function of separation in close pairs. The BPT catalogue categorises the close and wide pairs samples into Starburst, Transition, LINER and Seyfert classifications. LINER galaxies can show similar line widths to the narrow-line region in Seyferts and are often difficult to distinguish between in the BPT diagram (Heckman 1980). We define a LINER/Seyfert class to represent galaxies that have log([N\Rmnum{2}]/H$\alpha$) $> -0.2$ but have [O\Rmnum{3}] or H$\beta$ lines with signal-to-noise $\leq$3 and so can not be classified distinctly between LINER and Seyfert. We analyse how the Transition, LINER, Seyfert and LINER/Seyfert fractions change as pairs advance to the lowest separation bin. Our aim is to see if the AGN fraction rises; i.e. if AGN activity is somehow ignited in some close pairs as the merging process advances. The pairs sample is split into four projected separation bins (0-15kpc, 15-30kpc, 30-80kpc and 80-150kpc).

\begin{figure*}
\begin{center}
\begin{tabular}{cc}
\includegraphics[width=0.5\textwidth]{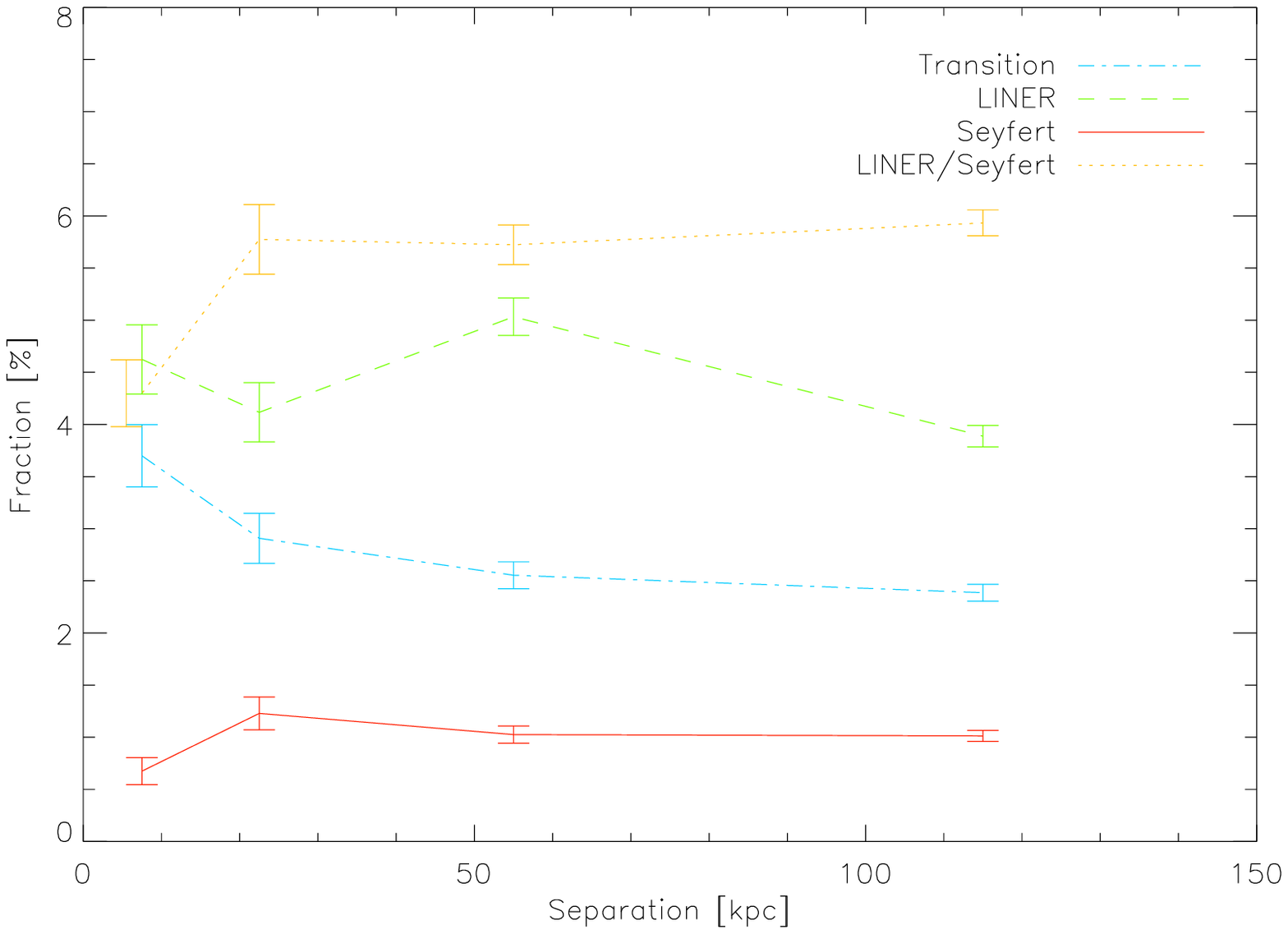} & \includegraphics[width=0.5\textwidth]{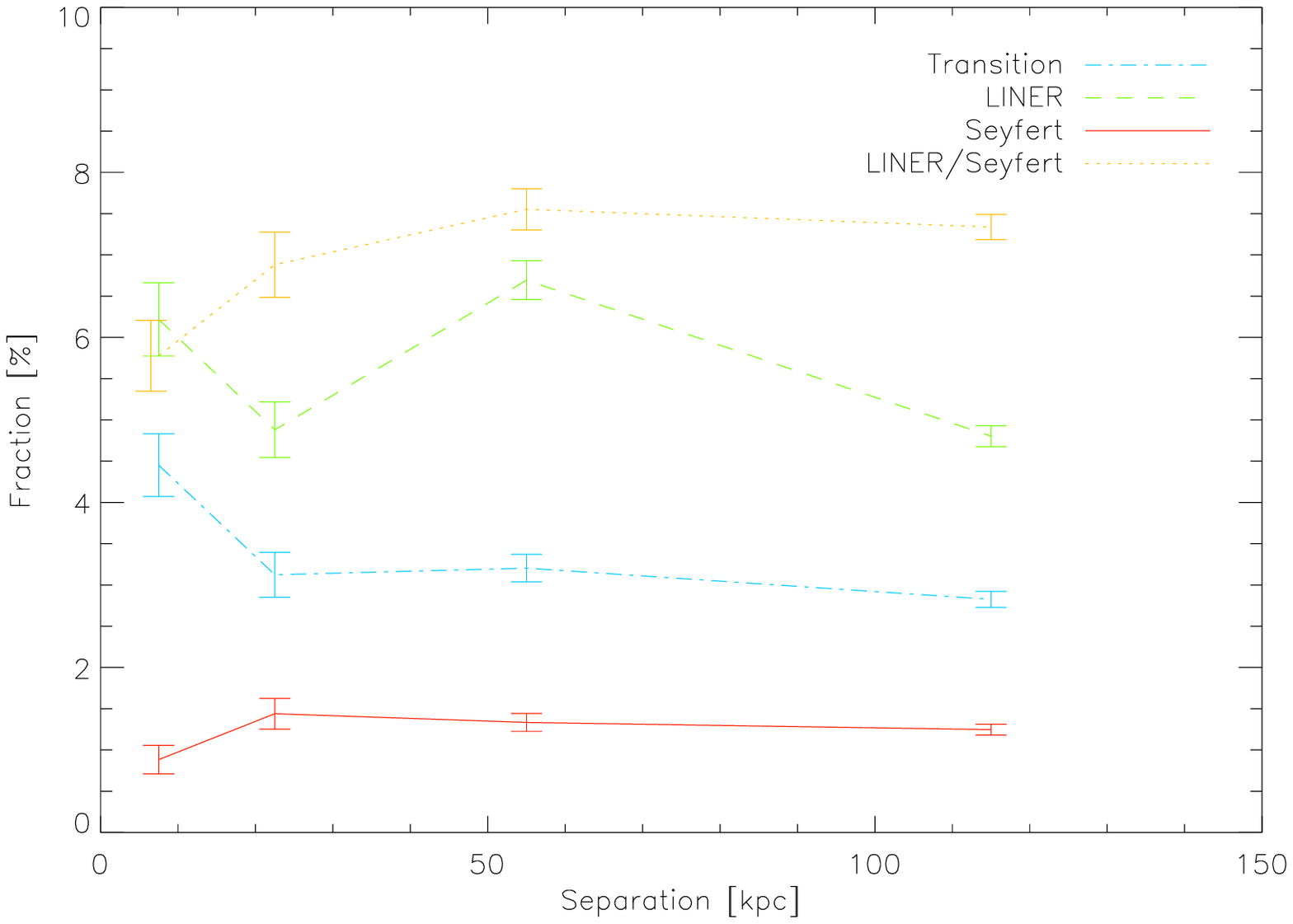} \\
\includegraphics[width=0.5\textwidth]{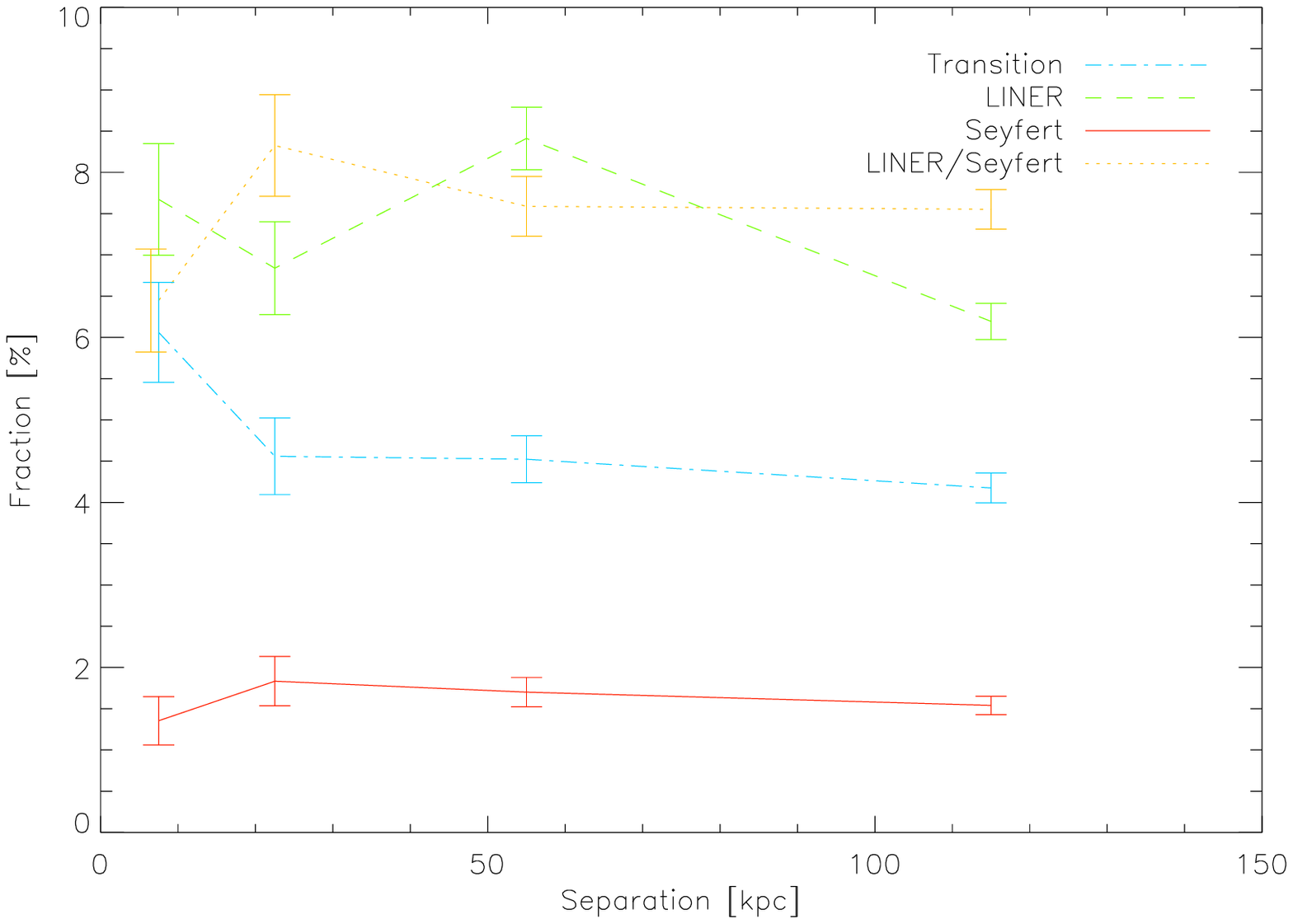} & \includegraphics[width=0.5\textwidth]{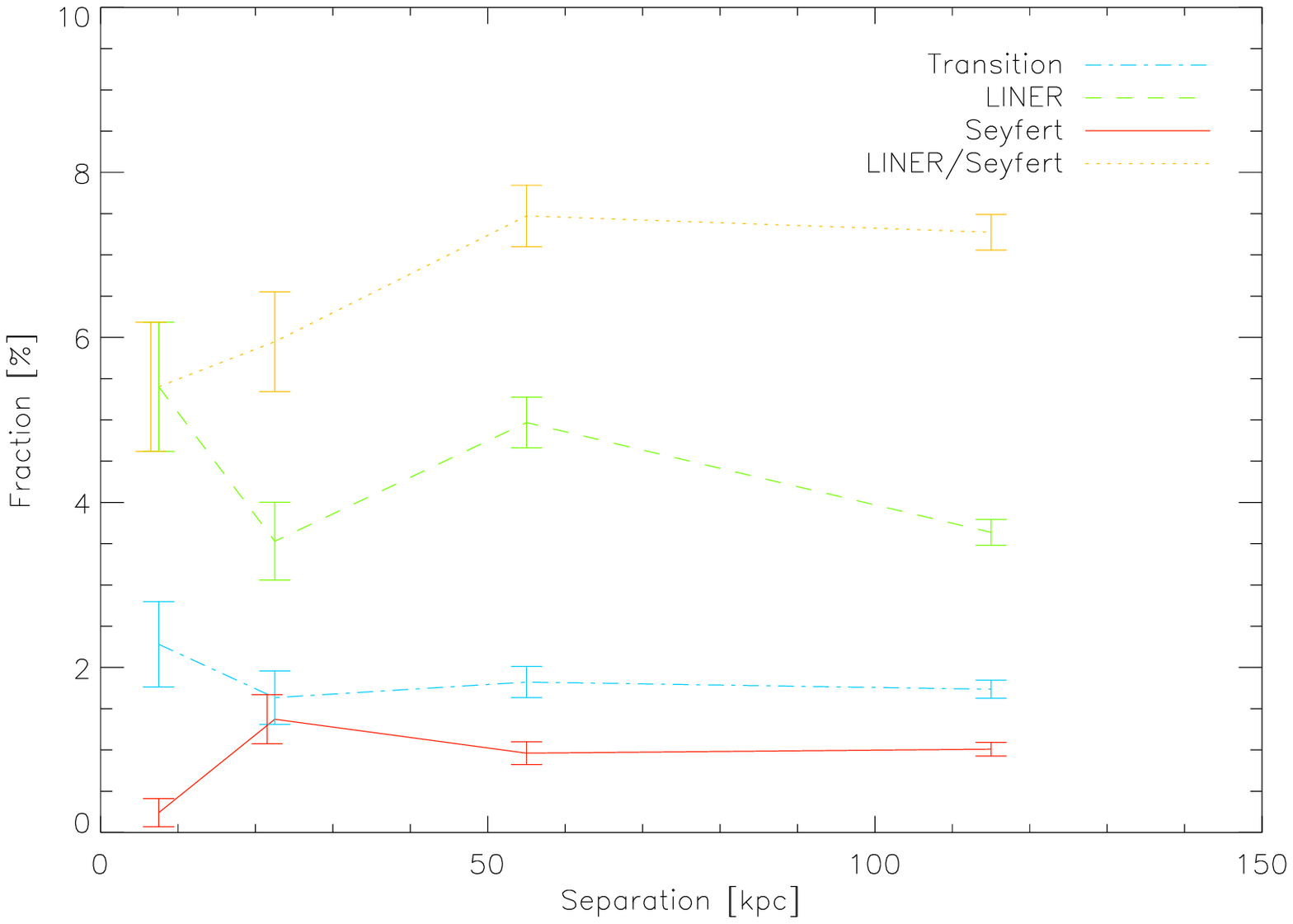} 
\end{tabular}
\caption{\small{Fraction of Transition, LINER, Seyfert and LINER/Seyfert galaxies in each of the following four separation bins: 0-15kpc, 15-30kpc, 30-80kpc and 80-150kpc. Top Left: All masses and environments. Top Right: Restricted to high mass (M$\geq$10$^{10}$M$_{\odot}$) since AGN activity is unlikely in lower mass galaxies.  Bottom left: High mass galaxies (M$\geq$10$^{10}$M$_{\odot}$) in field environments. Bottom right: High mass galaxies (M$\geq$10$^{10}$M$_{\odot}$) in group environments.}}\label{merger_evolution}
\end{center}
\end{figure*}

Figure \ref{merger_evolution} (top left) shows the fraction of galaxies in each separation and BPT classification bin. Although we have restricted our analysis to not include Starburst and Quiescent galaxies, we include these in the total sample when calculating fractions. The error bars shown are standard Poisson number count errors. As pair separation decreases, the fraction in the LINER/Seyfert class decreases significantly and we see a steady rise in the Transition class. We see a drop in the Seyfert fraction from 5.9\% (in the widest separation bin) to 4.3\% (in the smallest separation bin); i.e. a factor of 1.4 drop. Since galaxies with stellar mass below $\sim$$10^{10}$M$_{\odot}$ are unlikely to host AGN \citep{Kauffmann2003}, we now restrict the analysis to higher stellar mass galaxies (M$\geq$10$^{10}$M$_{\odot}$); see the top right plot in Figure \ref{merger_evolution}. This plot shows a similar distribution to the full stellar mass sample, but now each category accounts for a higher fraction. Here, we see a drop in the Seyfert fraction from 7.3\% to 5.8\%; a factor of 1.3 decrease.

We further split the sample into pairs in the field (Figure \ref{merger_evolution} bottom left) and pairs in group/cluster environments (bottom right). In the field, we see no significant increase or decrease in the Seyfert fraction, however, we do see a rise in the Transition fraction for the closest pairs. In high density environment pairs we see a significant drop in the Seyfert fraction from 1.0\% to 0.2\%; a factor of 4.2 decrease. Note that after splitting by stellar mass and environment, we are left with few galaxies in the smallest separation Seyfert bins. Note also that the LINER and LINER/Seyfert fractions are equal in the final plot; both bins contain only 45 galaxies. We are generally seeing a small decrease in Seyfert fraction in pairs at very low separation, paralleled with an increase in the Transition fraction. This suggests that AGN activity may increase but it may be overwhelmed by star formation in low separation close pairs.

\section{Discussion and Summary}
We study close pair galaxies using NUV-$r$ colours and NUV-derived specific star formation rates. This serves to add to previous work on close pairs in which optical colours and emission lines are used as indicators of star formation. Whereas emission line measurements are limited by finite fibre size and rely on corrections, NUV photometry provides a broader measure of young star formation for an entire galaxy. Our sample consists of SDSS optical spectroscopy and photometry, and GALEX photometry for low redshift close pair systems, and we also extract a sample of wide pair galaxies to use as a control sample representative of the general population. 

We find a factor of 5.3 increase (6.3$\times$10$^{-11}$yr$^{-1}$) in SSFR for low stellar mass close pair galaxies and a factor of 2.1 increase (1.3$\times$10$^{-12}$yr$^{-1}$) in SSFR for high stellar mass close pairs compared to the general galaxy population. We find a difference of 1.4$\times$10$^{-11}$yr$^{-1}$ (i.e. a factor of 1.8 increase) in SSFR from the widest to the smallest separation bins for pairs in field environments, and no significant increase for pairs in group and cluster environments. In the low mass sample we find a rise in SSFR of 4.4$\times10^{-11}$yr$^{-1}$ (a factor of 2.4 increase) for pairs in the field and a rise of 1.2$\times10^{-11}$yr$^{-1}$ (a factor of 3.4 increase) for pairs in groups. For high mass pairs we see a rise of 3.9$\times10^{-12}$yr$^{-1}$ (a factor of 2.5 increase) in SSFR in field environments.

These results are consistent with \cite{Wong2011}, who used NUV-r and FUV-r colours from GALEX as a proxy for SSFR for intermediate redshift close pairs ($0.25\leq z \leq0.75$) drawn from PRIMUS. They found an $\sim$$15-20\%$ increase in SSFR for close pairs with projected separation $\leq$50$h^{-1}$ kpc, and an $\sim$$25-30\%$ increase in SSFR for close pairs with projected separation $\leq$30$h^{-1}$ kpc. PRIMUS spectra is lower resolution than SDSS spectra (redshifts are accurate to within $\sigma_{z}/(1+z)$ with $\lesssim$3$\%$ catastrophic outliers), and the coverage is less (9.1 deg$^{2}$ of the sky, using multiple independent fields). Our SSFRs were calculated directly from the NUV luminosity, whereas Wong et al.\ use colours to estimate SSFR; yet despite different methods for estimating SSFR, the implications of enhancement in close pairs are consistent.  A combination of these two studies shows evidence for SSFR enhancement from local to intermediate redshift close pairs.

Our results are also consistent with previous work on close pairs using optical colours and emission lines as tracers of star formation. \cite{Woods2007} also find an enhancement in star formation in low mass galaxy close pairs using H$\alpha$ as a diagnostic for star formation. Using asymmetry effects and optical colours as tracers, \cite{Kauffmann2004}, \cite{Alonso2005} and \cite{Ellison2010} also find an enhancement in star formation in low density environments. This is likely to be due to the higher gas fraction available in low density environments to fuel star formation, since tidal fields and ram pressure stripping can lower gas fractions in higher density environments. 

We find a significant drop in the Seyfert fraction as interactions progress to the nearly coalesced stage. For all masses and environments, we see a factor of 1.4 decrease in Seyfert fraction from general population galaxies to close pairs. This is particularly prominent in high mass close pairs in group environments, where we find a factor of 4.2 decrease in Seyfert fraction. A steady rise is also seen in the fraction of galaxies in the Transition region of the BPT diagram.

We see strong evidence that merging can cause a change in emission line processes, leading to an evolution in a galaxy's location in the BPT diagram. However, based on our BPT analysis, where Seyferts are the only category to definitely harbour AGN activity, we see little evidence that mergers are triggering AGN activity during the close pairs stage of merging. We propose that, if AGN activity is ignited in some interacting massive galaxies as theoretically predicted, this process may lead to another class of AGN activity, or take place at the post-merger stage once the merging black holes have coalesced \citep[see][]{Carpineti2012}.

\section*{Acknowledgements}

We are grateful to the anonymous referee for many useful comments that improved the original manuscript. We thank Sara Ellison and David Patton for their helpful and constructive comments, Daniel Mortlock for his help with statistical calculations, and Mike Blanton for his publically available Kcorrect code. SK acknowledges research fellowships from the 1851 Royal Commission, Imperial College London and Worcester College Oxford and support from the BIPAC institute at Oxford.

Funding for the SDSS and SDSS-II has been provided by the Alfred P. Sloan Foundation, the Participating Institutions, the National Science Foundation, the U.S. Department of Energy, the National Aeronautics and Space Administration, the Japanese Monbukagakusho, the Max Planck Society, and the Higher Education Funding Council for England. The SDSS Web Site is http://www.sdss.org/. The SDSS is managed by the Astrophysical Research Consortium for the Participating Institutions. The Participating Institutions are the American Museum of Natural History, Astrophysical Institute Potsdam, University of Basel, University of Cambridge, Case Western Reserve University, University of Chicago, Drexel University, Fermilab, the Institute for Advanced Study, the Japan Participation Group, Johns Hopkins University, the Joint Institute for Nuclear Astrophysics, the Kavli Institute for Particle Astrophysics and Cosmology, the Korean Scientist Group, the Chinese Academy of Sciences (LAMOST), Los Alamos National Laboratory, the Max-Planck-Institute for Astronomy (MPIA), the Max-Planck-Institute for Astrophysics (MPA), New Mexico State University, Ohio State University, University of Pittsburgh, University of Portsmouth, Princeton University, the United States Naval Observatory, and the University of Washington.

GALEX is a NASA Small Explorer, launched in April 2003. We gratefully acknowledge NASA's support for construction, operation, and science analysis for the GALEX mission, developed in cooperation with the Centre National d'Etudes Spatiales of France and the Korean Ministry of Science and Technology.

\bibliographystyle{mn2e}
\bibliography{mn-jour,cp_refs_nobrackets}

\label{lastpage}

\end{document}